\documentclass[aps,prb,reprint,groupedaddress,twocolumn,10pt]{revtex4-1}
\usepackage[seperr, load={}]{siunitx} 
\usepackage{array,multirow,graphicx}
\usepackage{color}
\usepackage{xcolor, soul}
\sethlcolor{yellow}

\usepackage{textcomp}

\newcommand{\Vol}{\mathcal{V}}
\newcommand{\Qep}{\dot{Q}_\mathrm{ep}}
\newcommand{\Qnis}{\dot{Q}_\mathrm{nis}}

\newcommand{\Qtot}{\dot{Q}}
\newcommand{\Qs}{\dot{Q}_\mathrm{S}}
\newcommand{\Qn}{\dot{Q}_{0}}

\newcommand{\ns}{n_{\mathrm{S}}}
\newcommand{\ks}{\kappa_{\mathrm{S}}}
\newcommand{\pn}{\rho_{\mathrm{N}}}

\newcommand{\Tb}{T}
\newcommand{\Tp}{T}
\newcommand{\Te}{T_{\mathrm{e}}}
\newcommand{\TS}{T_{\mathrm{S}}}

\newcommand{\Gth}{G_\mathrm{th}}

\newcommand{\Ce}{\mathcal{C}_{\mathrm{e}}}

\newcommand{\kB}{k_{\mathrm{B}}}
\newcommand{\kF}{k_{\mathrm{F}}}
\newcommand{\kth}{\kappa_{\mathrm{th}}}

\newcommand{\Ph}{P_{\mathrm{H}}}
\newcommand{\Vh}{V_{\mathrm{H}}}

\newcommand{\Vb}{V_\mathrm{b}}
\newcommand{\Rt}{R_\mathrm{T}}

\begin{document}


\title{Anomalous electronic heat capacity of copper nanowires at sub-kelvin temperatures}

\author{K. L. Viisanen}
\author{J. P. Pekola}


\affiliation{Low Temperature Laboratory, Department of Applied Physics, Aalto University School of Science, P.O. Box 13500, 00076 AALTO, Finland}


\date{\today}

\begin{abstract}

We have measured the electronic heat capacity of thin film nanowires of copper and silver at temperatures 0.1–--0.3 K; the films were deposited by standard electron-beam evaporation. The specific heat of the Ag films of sub-100 nm thickness agrees with the bulk value and the free-electron estimate, whereas that of similar Cu films exceeds the corresponding reference values by one order of magnitude. The origin of the anomalously high heat capacity of copper films remains unknown for the moment. Based on the small heat capacity at low temperatures
and the possibility to devise a tunnel probe thermometer on it, metal films form a promising absorber material, e.g., for micro-wave photon calorimetry.

\end{abstract}

\pacs{}

\maketitle


%

  \section{Introduction}

  For improving the energy resolution
  of cryogenic bolometers and calorimeters,
  the thermal properties of
  mesoscopic structures play a key role.
  They can deviate significantly from the
  properties of bulk materials due to the
  surface effects that manifest in structures
  with a large surface-to-volume ratio.
  Breaking the periodic condition of a bulk
  material induces surface states
  and affects the bonding in the material, which
  can dramatically change its electric
  and magnetic properties by
  transforming conductors into insulators or
  semiconductors, and by
  inducing magnetism in materials that are
  non-magnetic in bulk. 
  Anomalous behavior can also
  result from impurities, dislocations in the lattice,
  or from a natural oxide layer growing
  on the surface.
  Due to its good thermal,
  electrical and mechanical
  properties, copper is widely used
  in low temperature experiments for instance in
  thermometry, calorimetry and 
  bolometry, development of electric
  current standards,
  refrigeration by nuclear
  demagnetization and
  electronic cooling \cite{Giazotto:2006_thermometry,Enss:2015_noise_thermometry,Goldie:2011_MoCu_TES,Viisanen:2015_relax,Viisanen:2015_calorim,Pekola:2013_current_standard,Todoshenko:2014_nuclear_demagnetization_cooling,
  ullomNIS:2013,Nguyen:cooler_2015}.
  
   \begin{figure}[h!]
 \centering
 \includegraphics[width = 0.48  \textwidth]{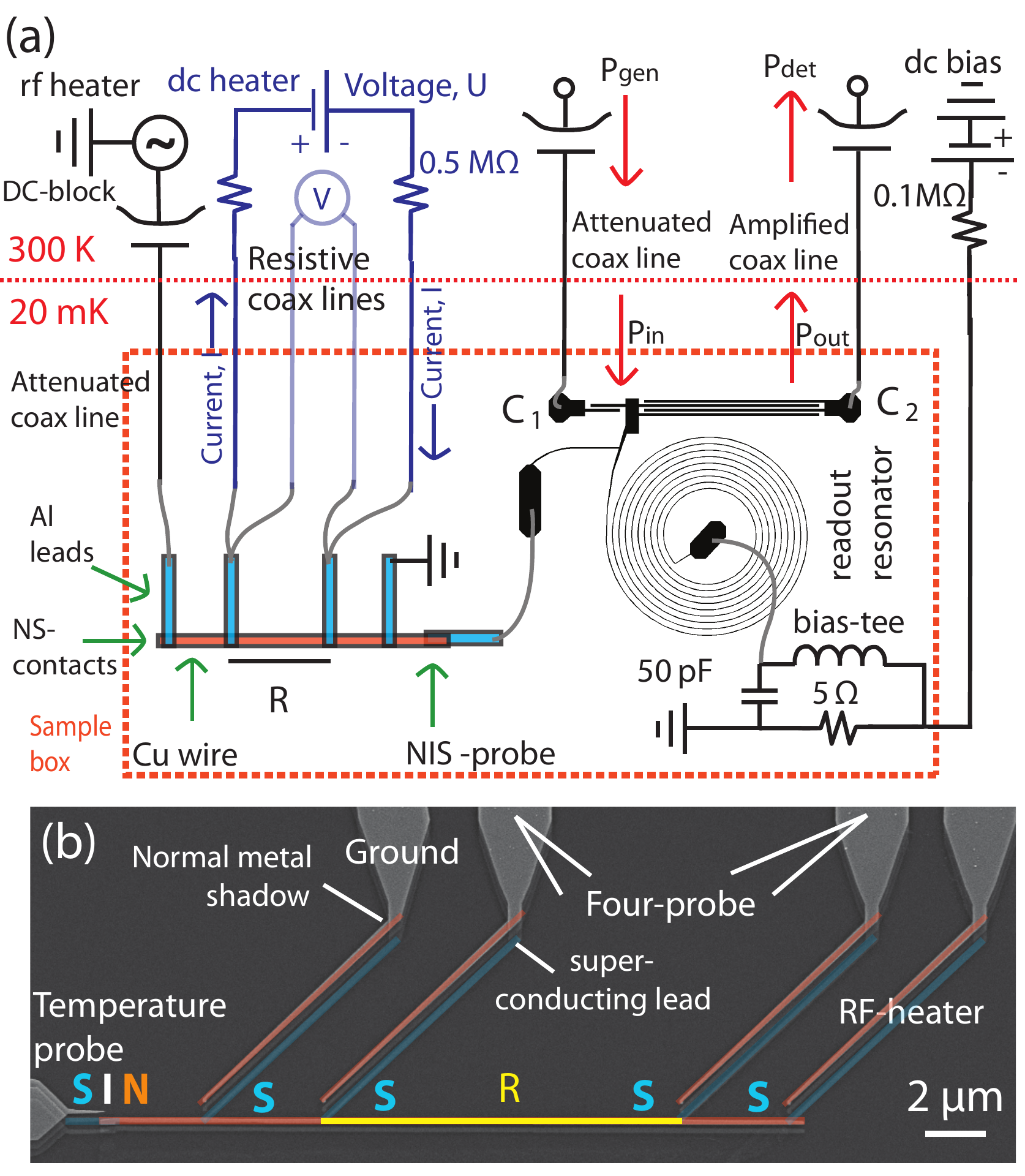}
 \caption{The set-up of the measurements. (a) A schematic presentation. Four coaxial direct current (dc) wires are connected to the sample for calibrating the heating power. A dc bias is applied to the NIS-temperature probe through a bias-tee and a cold bias resistor at the printed circuit board (pcb) of the sample box. The thermometer is operated with an rf-drive, directed to the sample through a wide band coaxial wire. The readout is performed through a lumped element Al resonator, which is operated in a transmission mode. The coupling to the $Z_0 = \SI{50}{\Omega}$ impedance
transmission lines is realized with
finger capacitors of capacitance
$C_1 = $\SI{0.025}{pF} and $C_2 = $\SI{0.22}{pF}.
The inductance of the spiral coil is \SI{94}{nH}
and the parasitic capacitance to ground from the resonator and the
sample is \SI{0.61}{pF}. 
Another rf-wire is connected to the sample to allow for the application of fast heating pulses. (b) Coloured micrograph of the sample for the heat capacity measurement. The resistance, $R$, is measured for the \SI{12}{\mu m} long section of the actual wire, painted in yellow.}
 \label{figure_setup}
 \end{figure}

  For applications in calorimetry, the main
  characteristics of metals
  under interest are, in addition to
  the electric properties,
  the specific heat of the conduction electrons
  and their thermal coupling
  to the environment.
  Magnetic impurities of per mil concentration level
  are observed to enhance the specific heat of Cu
  by a few times at sub \SI{1}{K}
  temperatures \cite{cryo:book} and
  the surface of Cu is proposed to host Kondo
  impurities \cite{Pierre:2003}
  in the dephasing time measurements,
  which are sensitive to even a dilute magnetic
  impurity concentration,
  intensively studied in mesoscopic Cu structures
  during the last decades
  \cite{Vranken:1988_dephasing,Pierre:2003}.
  The specific heat of an electron gas interacting with
  low concentration of randomly distributed magnetic
  impurities is determined by the Kondo temperature,
  $T_\mathrm{K}$, in the material and can exceed the
  free electron specific heat by orders of
  magnitude\cite{magnetic_impurities:2012}.

  The surface of Cu structures
  is usually covered by
  its natural oxides. $\mathrm{CuO}$ is paramagnetic
  even in bulk and $\mathrm{CuO_2}$
  has been observed to exhibit various magnetic properties
  such as antiferromagnetism and
  ferromagnetism in
  nanoparticles,
  even though the bulk structure is
  diamagnetic \cite{Ahmad:2005_CuOx_nanoparticles,Chen:2009_CuOx_nanoparticles}.
  Theoretical analysis of
  Cu oxide surfaces traces the origin
  of ferromagnetism in the pure
  $\mathrm{CuO_2}$ nanoparticles to the
  increased 2p-3d hybridization in the
  nanomaterial,
  and the modelling of cation vacancies in
  an ideal $\mathrm{CuO_2}$ crystal as well as on its
  surfaces indicates the
  vacancies to be a possible
  source of the observed magnetic
  moments \cite{Yu:2015_dft_CuO_surface,Chen:2009_CuOx_nanoparticles}.
  Cupric oxide layers play an important
  role also in the observed high transition temperature
  superconductivity of several heavy fermion compounds,
  where the effective mass can exceed that of
  bare electrons by two orders of magnitude
  resulting in significant increase
  in the specific heat of the
  compound \cite{Coleman:2007_heavy_fermions_book}.
  
  Lattice dislocations
  induce electric field gradients (EFG)
  in a material.
  Additional heat capacity arising from
  the nuclear spin coupling to the
  EFG created by
  crystal field distortions and magnetic impurities
  in metals has been discussed
  in Ref.~\cite{Siemensmeyer:1992}.
  Previous measurements in magnetic
  metallic calorimeters suggest that
  additional specific heat in Au arises
  from the nuclear quadrupole coupling
  to the EFG introduced by Er ions\cite{Enss:2000_calorimeter}.
  The accurate estimation of strain induced EFGs
  is in practice challenging, since each individual
  dislocation in the lattice should be
  included in the model.

  \section{Description of the experiment}
  
  In this paper we present
  specific heat measurements at temperatures
  $\SI{120}{}-\SI{250}{mK}$ of Ag and Cu thin film wires.
  The normal metal specific heat
  at low temperatures is given by
  $c = \gamma T + \beta T^3$,
  where the first term
  is due to the conduction
  electrons and the second one comes
  from the lattice phonons.
  At sub-kelvin temperatures, the
  phonons are frozen out and the
  electrons are left as the main source of
  heat capacity, whereas
  at higher temperatures, the electronic
  contribution is negligible.
  The specific heat measurements on bulk Cu and Ag 
  address the sum of the two
  components
  \cite{Martin:specific_heat:1968,Corak:atomic_heat:1955,kittel, pobell}.
  The literature values of the low temperature
  specific heat
  for bulk Cu are usually
  \SI{30}{\percent} higher than
  the free-electron estimate, 
  while Ag seems to be a manifestation
  of the free-electron
  theory.
  In those measurements,
  the electronic contribution is
  extracted based on the temperature
  dependence, whereas, in our setup,
  the electronic heat capacity
  is measured directly
  as we Joule heat the electrons by electric current.
  We observe that Cu wires exhibit an
  anomalously large specific heat, 
  exceeding the free-electron estimate
  by up to an order of magnitude, whereas
  the Ag wire follows the free-electron estimate.

\begin{table}[h]
\centering
\caption{Parameters of the samples and the resonators. The resistivities, $\rho$, of the metals are calculated from the measured $R$ and the dimensions of the wires, given in Table~\ref{table_sample_volume}.  }
  \begin{tabular}{ |  c |  c |  c |  c |  c |  c | c |c |c |c |}
    \hline
     Sample & Mate- &  $R$ & $\rho$& $R_\mathrm{T}$ & $\Delta$ & $R_0$ & $\omega_0/2\pi$ \\ 
     &rial &  $(\Omega)$ & $(\Omega m)$ & ($\mathrm{k\Omega}$) & $(\mathrm{meV})$ & ($\mathrm{k\Omega}$) & $(\mathrm{MHz})$  \\ \hline
    A & Cu & 96.9  & \SI{3.3e-8}{} & 29 & 2.16 & 34 & 550  \\ 
    B & Cu & 18.6  & \SI{3.0e-8}{} & 21 & 2.38 & 33 & 563  \\ 
    C & Ag & 80.5 & \SI{3.2e-8}{} & $11$ & 2.12 & 45 & 552  \\
  \hline
  \end{tabular}
\label{table_NIS}
\end{table}

\begin{figure}[b]
\centering
\includegraphics[width = 8cm]{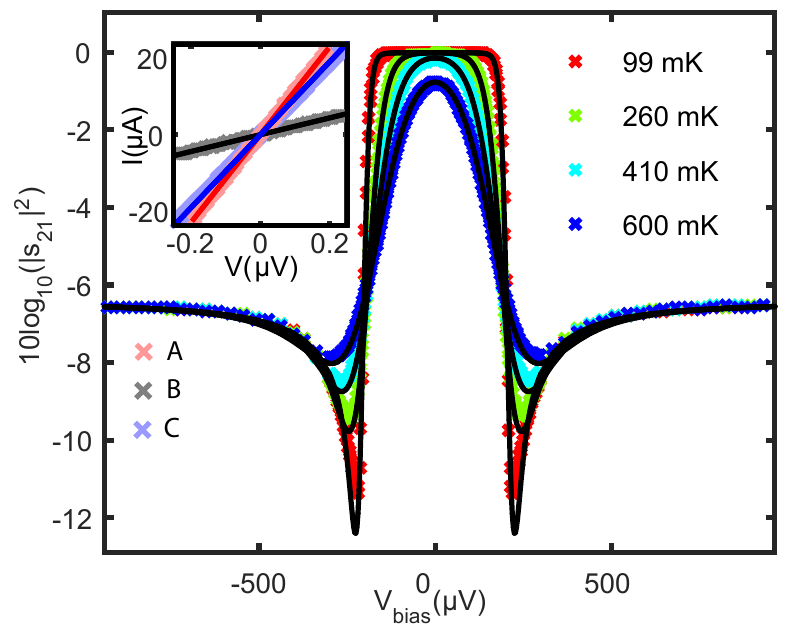}
\caption{Measured $|s_{21}|^2$ as a function of $\Vb$ in sample A. The theoretical estimates (black lines) are obtained with Eqs.~(\ref{s21}) and (\ref{nis_conductance}), assuming $\Te$ to be independent of $\Vb$. This is a reasonable asumption for samples with large normal metal volume at biases $e\Vb < \Delta$. In the top-panel, the current-voltage characteristics of the normal metal wires in samples A, B and C are shown. The solid lines are linear fits to the measured data. }
\label{resistances_rfiv}
\end{figure}


The thin film samples were patterned
by electron beam lithography and the
metals were deposited using standard
electron beam evaporation. We measure
the temperature of the wires using
a normal metal-insulator-superconductor
(NIS) probe embedded in a lumped
element tank circuit
  \cite{Rowell:1976_nis,Schmidt:2003_thermometry,Viisanen:2015_relax,Viisanen:2015_calorim}. A schematic illustration
  of the measurement setup and an image of the actual structure are shown in
  Fig.~\ref{figure_setup}.
  The thermometer can detect temporal evolution
  of temperature down to
  $\mathrm{\mu s}$ time resolution in mesoscopic metal structures,
  allowing the measurements of thermal relaxation rates
  and electronic heat capacities. In most cases the probe measures quasiparticle current of the junction \cite{Viisanen:2015_relax}, but when placed near a superconductor, it can also operate as a Josephson thermometer \cite{Saira:2016_supercurrent} with lower dissipation.
  Upon optimization, the device is a promising candidate
  for microwave calorimetry,
  in which case the finite volume of the
  electron gas in the normal wire works as an absorber
  for incoming radiation.
  For a sensitive calorimeter, it is necessary
  to minimize the size of the wire,
  whereas for the present heat capacity measurement,
  this is not essential.
  The NIS-junction is connected to a \SI{10}{MHz}
  bandwidth lumped element
  resonator with $\omega/2\pi = $\SI{0.5}{GHz}
  resonance frequency, and it is operated here
  in the temperature range of $\SI{80}{}-\SI{300}{mK}$.
  The heat capacity,
  $\Ce$, of the conduction electrons
  in the normal metal island is measured
  by heating the wire with
  a voltage pulse and probing the electron
  temperature, $\Te$,
  in the normal metal. The thermal response of the wire
  depends on the magnitude of the heating pulse,
  $\Ph$, the heat conduction from the normal metal
  electrons to the environment, $\Gth$, and $\Ce$.
  All these quantities are determined separately
  in our measurement setup.

  \begin{figure}[t]
 \centering
 \includegraphics[width = 0.48  \textwidth]{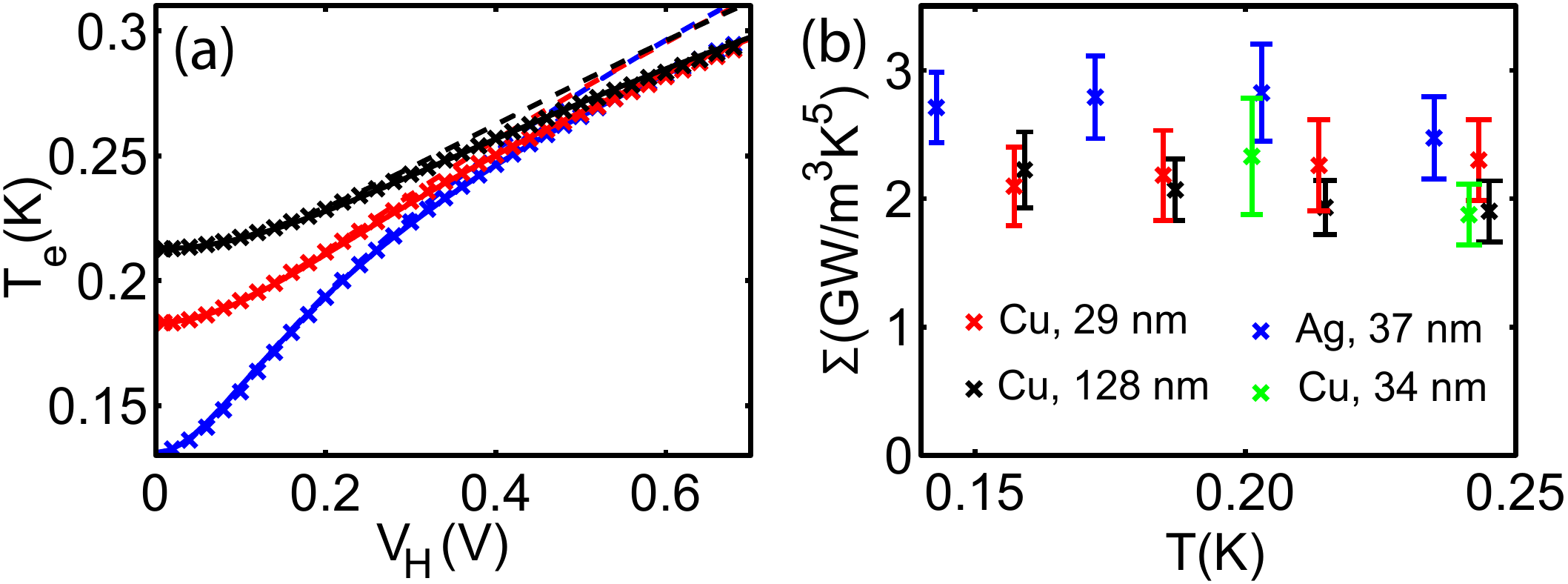}
 \caption{Measurements of thermal coupling of the electrons in the wire to the bath. (a) Measured steady-state temperature $\Te$ of sample A (28 nm thick Cu) against the amplitude of the continuously applied ac heating voltage at different bath temperatures. The solid lines are the theoretical estimates corresponding to $\Ph + \Qep+ \Qnis + \Qs = 0 $, $\Sigma$ and $\pn$ ($\approx \SI{2E-8}{\Omega m}$) are used as fitting parameters. The dashed lines are calculated by neglecting $\Qs$, which is justified at temperatures below \SI{250}{mK}. (b) The electron-phonon coupling constant, extracted for samples A-D, shown against bath temperature $T$.}
 \label{figure_samples}
 \end{figure}
 
  The heating power applied to the
  electron gas
  is determined in a four probe measurement, see Fig.~\ref{figure_setup}.
  Two pairs of dc lines
  are connected to the island through
  direct NS-contacts.
  The resistance
  of the section of the wire
  between these contacts, $R$,
  is measured by applying dc current,
  $I$, through it and
  measuring the voltage,
  $U$, across it.
  The measurement data is shown in
  the subfigure of
  Fig.~\ref{resistances_rfiv}. The linear slope in
  the $IU$-curve gives the resistance of the wire,
  see Table~\ref{table_NIS}.
  In the heat capacity measurement,
  the dc lines are left floating
  and the heating voltage, $\Vh$, is applied
  through a low-pass filtered and attenuated rf-line.
  The magnitude of the rf heating
  is given by $\Ph = A \Vh^2$, where
  the constant $A$ is determined by
  the attenuation in the heater line and the
  resistance of the normal wire.
  In practice, $A$ is determined
  by comparing the response
  of $\Te$ to the rf power and dc heating $UI$.

  At resonance, the transmittance of voltage in the circuit
  can be written as 
  \begin{equation}
  |s_{21}| = 2\kappa G_0/(G+G_0),
  \label{s21}
  \end{equation}where $\kappa$ is about $C_1/C_2$,
  $G_0 = Z_0(C_1^2+C_2^2)\omega_0^2$ and
  $G=R_{\mathrm{nis}}^{-1}$ is the
  differential conductance of the junction,
  given by
  \begin{equation}
  G = \frac{1}{\kB\Te\Rt}\int^{\infty}_{-\infty}{dE}\ns (E)f(E-e\Vb)(1-f(E-e\Vb)).
  \label{nis_conductance}
  \end{equation}Here $\Te$ is the electronic temperature in the normal metal, $f$ is the Fermi function, $f(E)~=~[1+\exp{(-E/{\kB\Te}})]^{-1}$ and
  $\ns~=~n_0 |\mathrm{Re} ( E/
  \sqrt{E^2-\Delta^2})|$ is the Bardeen-Cooper-Schrieffer density of states (DOS) of the superconductor, with
  $n_0$ the normal metal DOS at the Fermi level.
  The tunnelling resistances, $R_\mathrm{T}$,
  of the NIS-junctions
  are measured in room temperature.
  The parameters of the resonators and the
  junctions are shown in
  Table~\ref{table_NIS}.
  Typically, $R_\mathrm{T}$ increases
  by about \SI{10}{\percent}\cite{Gloos:2003}, when cooled
  down to sub-Kelvin temperatures.
  By measuring the $|s_{21}|^2$ vs
  $\Vb$ characteristics of the samples,
  one obtains an estimate for $R_0/\Rt$.
  To determine $R_0$, the values of $\Rt$
  measured in room temperature,
  increased by \SI{10}{\percent}, are used.  
  Fig.~\ref{resistances_rfiv}(b) shows
  the $|s_{21}|^2$ vs $\Vb$ characteristics
  measured in sample
  A at a few different bath temperatures.
  The superconducting gap is
  obtained by fitting to the data at base temperature.

 \section{Thermal conductance measurements}
  
 The steady-state temperature
  on the normal metal
  island is governed by the heat balance
  $\Qtot + \Ph =0$.
  In the present configuration, $\Qtot$ is given by $\Qtot=\Qep+\Qnis+\Qs+\Qn$,
  where $\Qep~=~\Sigma\Vol(\Tp^5-T_e^5)$ is
  the power to the electron gas of the metal wire via electron-phonon scattering,
  $\Qnis$ is the heat carried by the tunnelling quasiparticles
  across the NIS-temperature probe
  and $\Qs$ is the heat leak across the
  superconducting aluminium leads. Here, $\Sigma$ is a material-specific parameter to be discussed below,
  $\Vol$ is the volume of the normal metal island
  and $T$ is the temperature of the phonon bath.
  The background power from the environment to the island, $\Qn$,
  is assumed to be constant (at all bath temperatures), and we assume that $T$
  does not depend on the heating power.
  Due to $\Qn$,
  $\Te$ of the carefully thermally isolated nanowire saturates to a value close to \SI{100}{mK}
  even without external power. 
  Since all the leads directly connected to the normal
  metal wire are superconducting, the heat
  flow along them is suppressed at
  low temperatures.
  For a superconducting wire
  of length $l$, thickness $t$ and width $w$, $\Qs$
  can be estimated at temperatures $\kB\TS << \Delta$ with
  $\Qs \approx -\frac{wt}{l}\int \limits_{\TS}^{\Te} dT' \ks (T')\mathrm{e}^{-\Delta/\kB T'}
  $, where $\ks (T')= \frac{6}{\pi^2}\frac{\mathcal{L}\Delta}{\kB\pn} \frac{\Delta}{\kB T'}$, $\mathcal{L}$ is the
  Lorenz number and $\pn$ is the normal state resistivity
  of the superconductor.
  In the present sample, there are four superconducting
  leads directly connected to the normal metal wire,
  each of them thermalized to $\Te$
  at the NS interface on the wire and to $\Tb$
  at the intersection between the
  superconducting lead and the normal metal shadow.
  $\Qnis$ is given by $\Qnis=\frac{1}{e^2\Rt}\int\ns(E)(E-e\Vb )[f_{\mathrm{N}}(E-e\Vb)-f_{\mathrm{S}}(E)]\mathrm{d}E$. $\Rt$ and $\Vol$ of the measured samples
  are sufficiently large, such that $\Qnis \ll \Qep$, which allows us to extract
  $\Sigma$ in the normal wire by measuring $\Te$ under
  different heating powers, as
  shown in Fig.~\ref{figure_samples}(a).


  The electron-phonon coupling constant $\Sigma$ is
  measured in samples A-D. The results are shown
  as a function of $\Tb$ in
  Fig.~\ref{figure_samples}(b), and the
  average values are gathered in
  Table~\ref{table_sample_volume}.
  The parameters are determined by measuring
  the steady-state temperature under heating
  (samples A-C) and as a reference in a standard
  SINIS cooling experiment  \cite{Giazotto:2006_thermometry}
  using a DC measurement with four NIS tunnel junctions (sample D).
  The values measured in this work,
  $\Sigma \approx  \SI{2E9}{W/m^3K^5}$ for Cu 
  and $\Sigma \approx  \SI{3E9}{W/m^3K^5}$
  for Ag
  are in agreement with
  previously measured values for Cu \cite{Meschke_Cu_sigma:2004}.
  For Ag, there is less data found
  in literature, but an estimate of \SI{0.5E9}{W/m^3K^5}
  was inferred by using
  $\Sigma$ as a fitting parameter
  in an experiment of
  Ref. \cite{Steinbach_Ag_sigma:1996}.
  For reference, rough theoretical 
  estimates of $\Sigma$ can be written
  with the density, $\rho$, the speed
  of sound, $c$, and the Fermi
  wave vector,
  $\kF$, in the metal as
  $\Sigma = \frac{\zeta(5)}{3\pi ^2}\frac{\kF^4\kB^5}{\hbar^3c\rho}$
  \cite{Wellstood:1994_eprel},
  where $\zeta(5)\approx \SI{1.0369}{}$.
  For Ag and Cu, this would give values $\Sigma=\SI{4E8}{W/m^3K^5}$ and $\Sigma=\SI{2E8}{W/m^3K^5}$, respectively. 

  \begin{figure}[b]
 \centering
 \includegraphics[width=8.5cm]{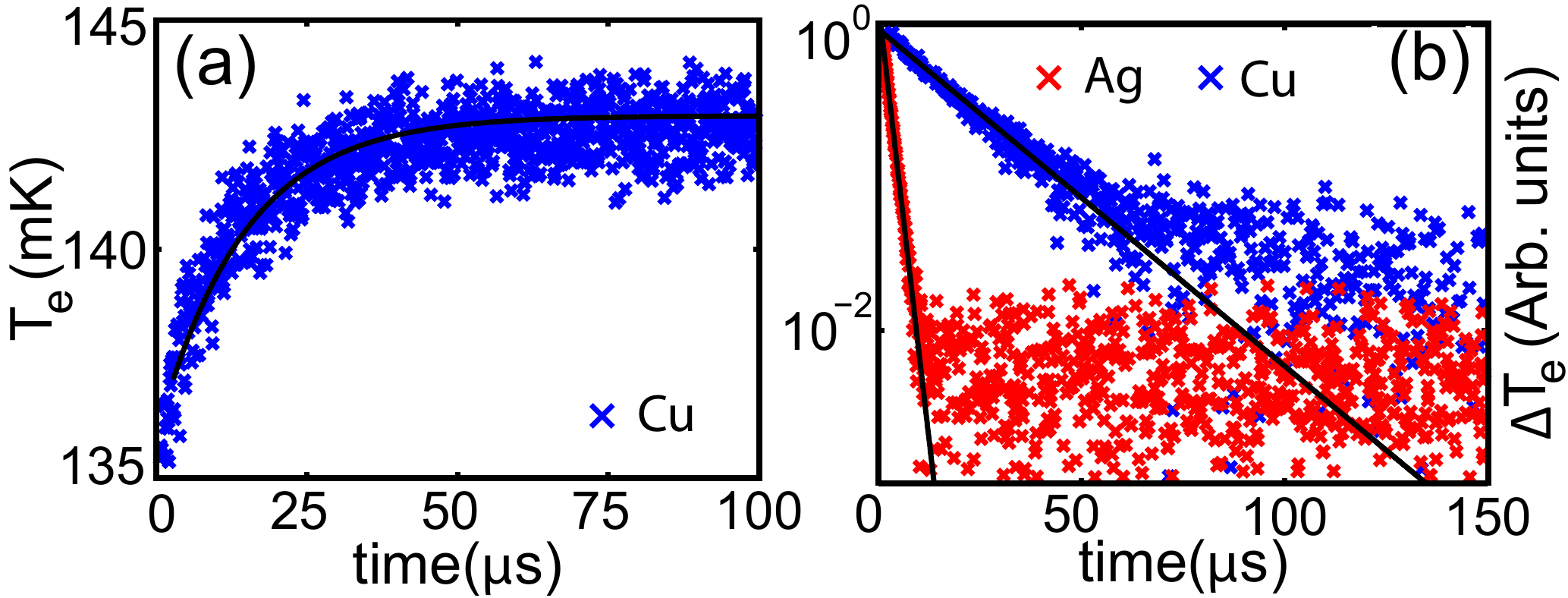}
 \caption{Thermal relaxation measurements. (a) Time evolution of $\Te$ in a thin Cu wire (sample A) after switching on the heating pulse. The black solid line is calculated by solving the heat balance equation~(\ref{eq_heat_balance}). (b) Thermal relaxation of thin Cu (sample A) wire at $T=107$ mK (blue) and Ag (sample C) wire at $T=102$ mK (red) after switching off the heating. The quantity on the vertical axis, obtained by subtracting the background and normalizing the signal, is proportional to $\Delta T_{\rm e}=T_{\rm e}-T$. The solid lines are exponential fits to the data. The timetraces are obtained by averaging over $10^5$ repetitions.}
 \label{figure_traces}
 \end{figure}

\begin{figure}[t]
\includegraphics[width=7cm]{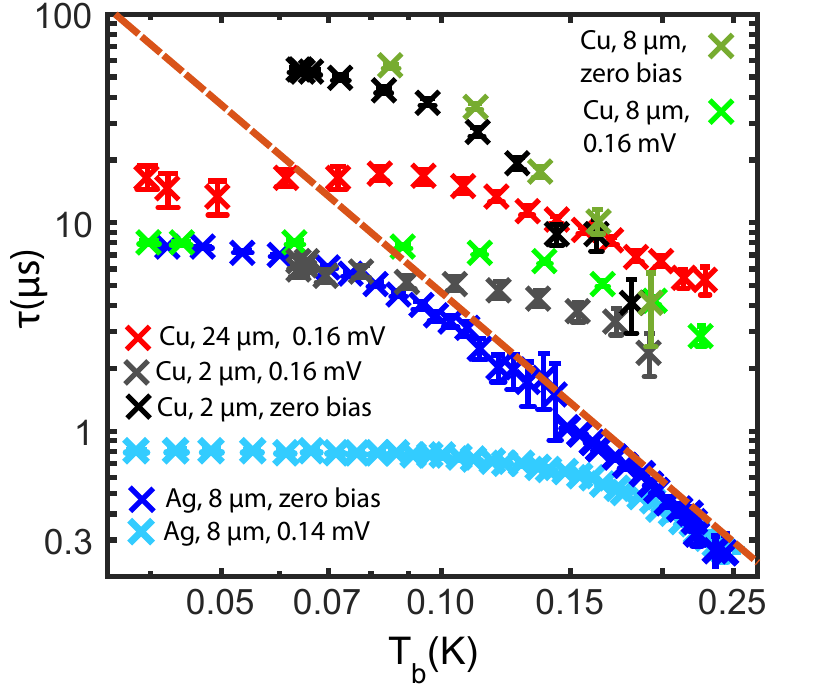}
\caption{Thermal relaxation time constants observed
in a silver wire and three copper wires of different lengths. The dashed line
shows the $\tau = \gamma/(5\Sigma T^{3})$ dependence, using the free-electron Sommerfeld coefficient $\gamma$ and the measured $\Sigma$ of Ag
($\gamma/(5\Sigma) = \SI{4.6E-9}{sK^3}$).
The decrease of the error bars in the Ag data at \SI{150}{mK} is due to the increased averaging and larger heating amplitudes used at higher temperatures. The data measured at zero-bias indicates slower thermal relaxation both in Cu and Ag. This can be partly due to the enhanced heat transport via quasiparticle tunnelling compared with the electron-phonon heat current in the smaller samples. More probable explanation is, however, the saturation of $\Te$ at low temperatures due to the noise induced by biasing the junction. }
\label{figure_relaxation_times}
\end{figure}

  \section{Heat capacity measurements}
 
  We measure $\Ce$ of a normal metal wire
  by heating it with a voltage pulse of finite length
  and observing the time-dependent temperature $\Te$
  of the electron system.
 Its time evolution 
  is determined by the heat balance equation
  \begin{equation}
  \Ce \frac{\mathrm{d} \Te}{\mathrm{d} t}= \Qtot + \Ph.
  \label{eq_heat_balance}
  \end{equation}
  The measured $\Te$
  after switching on a sinusoidal
  \SI{6}{MHz} heating pulse is shown in
  Fig.~\ref{figure_traces}(a) as a function of time.  
  Below $\Tb = \SI{250}{mK}$,
  due to the suppressed heat flow
  across the superconducting leads,
  the heat exchange between the normal metal electrons
  and the environment
  occurs mainly through the electron-phonon scattering
  as was demonstrated in Fig. \ref{figure_samples}. 	
  Since $\Sigma$ was measured independently in a DC
  measurement, $\Ce$ can then be determined accurately
  from the measured time traces.
  In particular, for small enough power, $\Te$ decays
  exponentially back to equilibrium
  with the time constant
  $\tau = \Ce/G_{\rm th}$ after switching off the heating pulse,
  see Fig.~\ref{figure_traces}(b).
  The thermal conductance is given
  by $G_{\rm th} = -\frac{d\dot Q_{\rm ep}}{dT_{\rm e}} = 5\Sigma\Vol\Te^4$.
  The heat capacity of the conduction electrons in
  a conventional metal can be estimated with
  the free-electron model as
  $\Ce = \gamma\Vol\Te$, where $\gamma= \frac{\pi^2}{3} n_0 k_B^2$
  is the Sommerfeld coefficient for the metal. For this ideal system,
  we thus expect the relaxation time to
  obey $\tau = \gamma/(5\Sigma\Te^3)$.

  We have measured the thermal relaxation times
  of Cu and Ag wires at different bath temperatures,
  $T$. The thermal relaxation
  times after switching off the heating are
  obtained by fitting an exponential function
  to the measured data, from which the background
  is subtracted and the trace is normalized.
  There is a $\sim 2$ $\micro$s electrical
  transient when switching the heating
  pulse on and off: data over this time
  is not included in the fits. In the
  Cu samples, the relaxation seems to
  consist of two time constants, whereas
  in the Ag samples,
  the relaxation can be fitted with a
  single exponent, $\tau$, which is clearly
  faster than the time constants in Cu.
  This is the most direct evidence of the
  largely different specific heats of Ag and Cu wires,
  see Fig.~\ref{figure_traces}(b).
  In Fig.~\ref{figure_relaxation_times},
  the thermal relaxation in Cu and Ag
  wires are shown as a function of $T$.
  The time constants shown for Cu are
  obtained with a two-exponent fit,
  while only the faster time constants
  are included in the plot.
  The slower relaxation observed
  in the Cu samples depends on the
  amplitude and the length of the
  heating pulse, increasing up to
  around \SI{150}{\micro s}
  at large heating voltages.
  This relaxation is possibly
  related to an additional heat capacity,
  that relaxes slowly after switching off
  the heating pulse. When the heating pulse
  is short and the temperature rise only a
  few mK, this effect is negligible.
  To rule out thermal gradients along the
  wire, we have measured the temperature
  relaxation in two additional Cu wires of
  lengths \SI{8}{\mu m} and \SI{2}{\mu m},
  see Fig.~\ref{figure_relaxation_times}.
  By describing the thermal relaxation of the
  free electrons in the normal metal wire
  with the Wiedemann-Franz relation between
  the thermal and electrical conductivities,
  $\kappa/\sigma = L\Te$, where $L$ is the Lorenz number,
  the response time of the electron temperature in the wire
  can be estimated as
  $\tau_{\mathrm{e-e}}= \Ce/\kth=\gamma\rho l^2/L$.
  For the \SI{24}{\mu m} wires this
  results in about \SI{50}{ns},
  which is about three decades smaller than the
  timescales observed in the Cu wires.
  Due to the quadratic dependence on length,
  the relaxation in the \SI{2}{\mu m} wire should be over
  two decades faster compared with the
  \SI{24}{\mu m} wires, which is not
  observed in our measurements.
  From these perspectives we can rule out
  the possibility of significant thermal gradients
  along the wires.

\begin{figure}[t]
 \centering
 \includegraphics[width=8cm]{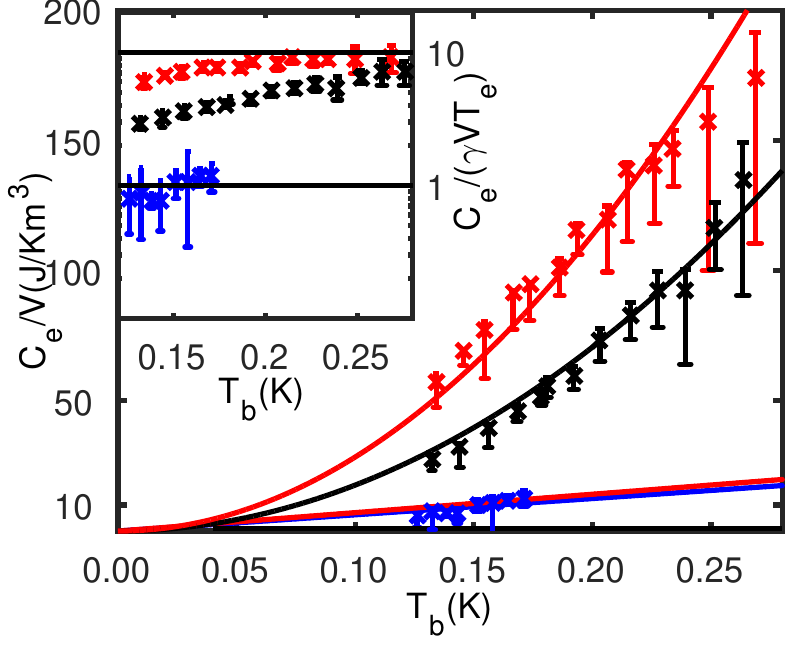}
 \caption{Specific heat of the normal metal wires measured at different bath temperatures. The solid lines correspond to the free-electron estimate for Cu (red) and Ag (blue). The data for the Ag wire falls close to the free-electron estimate, while the specific heat of the Cu wires is clearly higher with rather a quadratic dependence on $\Tb$ (dashed lines). The normalized specific heat, $\Ce/(\gamma\Vol T)$, is shown as a function of $\Tb$ in the inset of the figure. Here $\gamma$ is the Sommerfeld coefficient for each metal.}
 \label{figure_specific_heat}
 \end{figure}

  The specific heat of samples A-C, measured at different 
  bath temperatures,
  is shown in Fig.~\ref{figure_specific_heat}.
  The data are obtained by fitting the heat balance
  equation~(\ref{eq_heat_balance}) to the measured
  time evolution of $\Te$ after switching on a
  \SI{6}{MHz} sinusoidal heating voltage,
  as illustrated in Fig.~\ref{figure_traces}.
  Compared to the Cu wires, the thermal relaxation time of the
  Ag wire is clearly shorter, and hence its specific heat
  is lower. 
  The results are similar to those obtained from the decay traces
  after switching off the pulse.
  In practice the fast relaxation means that the measurement
  in Ag wire is limited to temperatures well below 0.2 K
  because of the finite bandwidth of the thermometer. 
  The largest specific heat is observed in the thin Cu wire.
  Especially with large amplitude
  heating pulses, an extra, very slow relaxation
  is observed in copper samples, which is evident in the
  non-exponential decay in Fig.~\ref{figure_traces}(b)
  as well. Therefore, for the analysis of copper data,
  only the beginning of the trace, $\sim 1.5\tau$ after
  switching on the heating pulse, was included in the fits.
  Within the free-electron model, the Sommerfeld coefficient $\gamma$
  obtains values $\gamma= \SI{70.7}{J/m^3K^2}$
  and $\gamma = \SI{62.4}{J/m^3K^2}$ for Cu and Ag, respectively. 
  According to the data in Fig.~\ref{figure_specific_heat},
  the Ag wire follows the free-electron theory,
  while an anomalously large
  specific heat is observed in the Cu wires.
  The specific heat of the thinner Cu
  wire exceeds the expectation
  value by an order of magnitude, while
  that of the thicker wire is somewhat lower.
  The temperature dependence in the two Cu
  samples is nearly quadratic, rather than linear
  predicted by the free-electron theory.

\begin{table}[b]
\centering
\caption{Physical dimensions and the average values of $\Sigma$ of the normal metal wires measured in this work. The resistances of the metals are measured for the \SI{12}{\mu m} long section of the $l=\SI{24}{\mu m}$ long wires. }
    \begin{tabular}{ |  c |  c |  c |  c |  c |  c | c | }
    \hline 
   \parbox[t]{3mm}{\multirow{3}{*}{\rotatebox[origin=c]{90}{Sample}}} &\parbox[t]{2mm}{\multirow{3}{*}{\rotatebox[origin=c]{90}{Material}}} & & & & & \\
     &  &$l$ & $w$ & $t$ & $\Vol$ & $\Sigma$\\ 
     & & ($\mathrm{\micro m}$) & $(\mathrm{nm})$ & $(\mathrm{nm})$ & $(\mathrm{\mu m^3})$ & $\mathrm{\Big(\frac{GW}{m^3K^5}}\Big)$  \\ \hline
     A & Cu & 24$\pm$0.1 & 145$\pm$5 & 28$\pm$1 & 0.10$\pm$0.01 & 2.2$\pm$0.2  \\ 
     B & Cu & 24$\pm$0.1 & 150$\pm$10 & 128$\pm$1 & 0.43$\pm$0.03 & 2.0$\pm$0.1\\ 
     C & Ag & 24$\pm$0.1 & 125$\pm$5 & 37$\pm$2 & 0.11$\pm$0.01 & 2.7$\pm$0.3 \\
     D & Cu & 7.2$\pm$0.1  & 180$\pm$10 & 34$\pm$1 & 0.044$\pm$0.004 & 2.1$\pm$0.3 \\
  \hline
  \end{tabular}
\label{table_sample_volume}
\end{table}

  \section{Detailed characterization of the samples}

  The measured samples are fabricated by
  electron-beam lithography and multi angle
  metal deposition in an electron-beam evaporator.
  In the evaporator,
  the samples are mounted on a
  room temperature platform in a
  vacuum chamber.
  During evaporation, the pressure in
  the chamber is between \SI{1E-7}{mbar}
  and \SI{1E-6}{mbar} and the
  growth rate of the film is
  in the range \SI{0.1}{}---\SI{0.3}{nm/s}.
  Typically the
  samples are stored in ambient air
  for a period ranging from
  hours to days before the
  cooldown to mK temperatures.
  We have observed, that the junction resistance of an
  $\mathrm{Ag-AlO_2-Al}$ tunnel contact
  increases when stored in room temperature.
  Indeed, in about a day, the electrodes
  detach completely.
  We have found this to be a
  typical problem with the Ag samples,
  while the $\mathrm{Cu-AlO_2-Al}$
  tunnel junctions are more
  robust.

   \begin{figure}[t]
  \includegraphics[width=8cm]{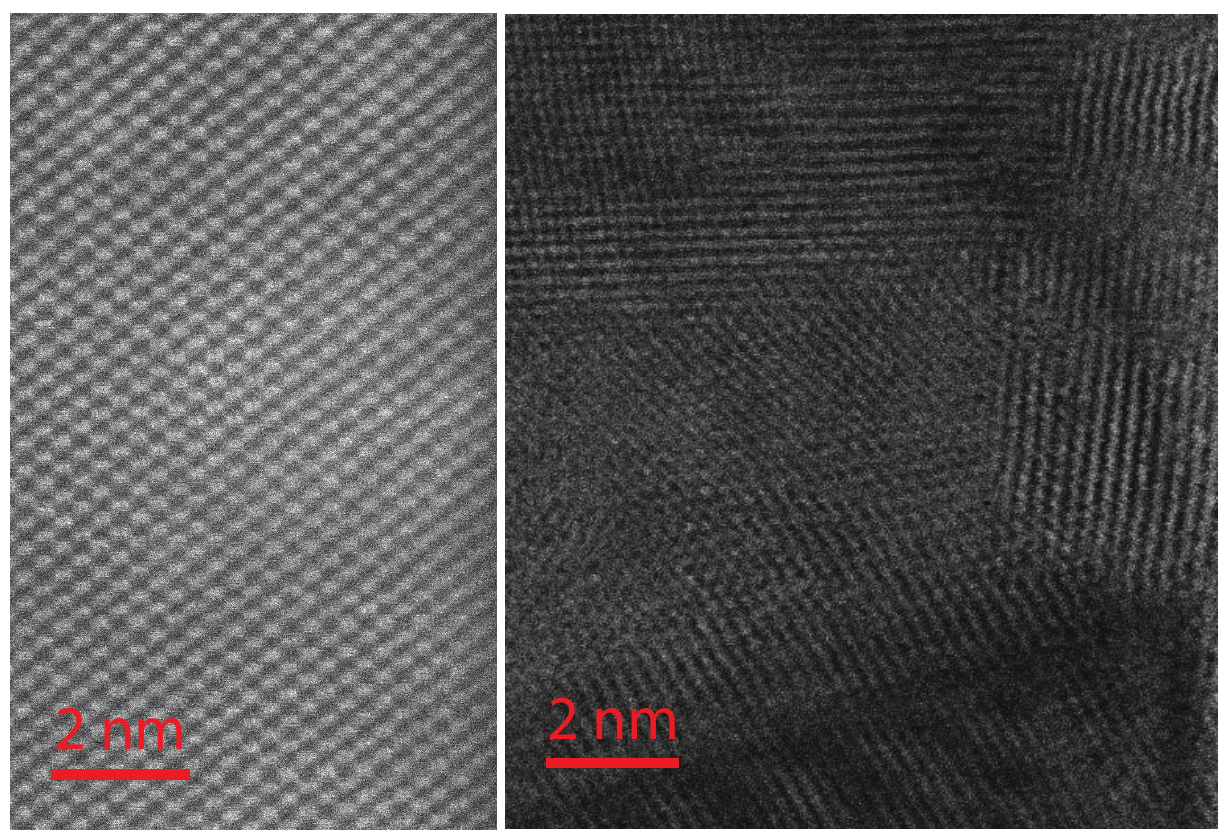}
\caption{STEM images of samples of (a) Ag and (b) Cu thin film nanowires. Several wires are imaged and all the Ag samples show a uniform lattice structure throughout the wire. The copper lattice, however, is formed of different growth directions, typical for evaporated Cu. }\label{figure_STEM}
\end{figure}
  
  The geometry and the atomic composition of
  the samples are analyzed with scanning
  electron microscopy (SEM),
  atomic force microscopy (AFM),
  secondary ion mass spectroscopy (SIMS)
  and scanning transmission electron microscopy (STEM).
  The length and width of
  the wires are determined by SEM imaging, while
  the thickness is measured by AFM.
  These values are shown in Table.~\ref{table_sample_volume}.

  The STEM imaging of the lattice
  structures is performed
  for \SI{100}{nm} thick cross-sections
  of the wires, cut out by milling with focused
  ion beam (FIB).
  Before exposing to FIB, the samples are
  covered with a protective layer of Au.
  Several wires are measured and the
  lattice structure is
  found to be similar to that of bulk, no
  gas pockets or other anomalies are
  observed. The images
  are shown in Fig.~\ref{figure_STEM}.
  The lattice structure of the
  evaporated Ag wires is
  remarkably uniform; the lattice
  vectors point to the same
  direction throughout the wire.
  In the Cu wires, different growth directions
  are observed, which is typical for
  evaporated Cu films.

  The amount of magnetic impurities in the
  evaporated metals is determined by
  SIMS, the level of \SI{0.6}{ppm} is measured in Cu and
  \SI{40}{ppm} in Ag \cite{eag}.
  The analysis is carried out for
  \SI{50}{nm} thick films, fabricated on a similar
  substrate and deposited in the same
  evaporator as the measured samples.
  The Cu film fabricated of 99.9999 \% pure source
  material was observed to be fairly clean with
  magnetic particles below the ppm level
  (Fe \SI{0.2}{ppm},
  Cr \SI{0.02}{ppm},
  Ni \SI{0.2}{ppm},
  V \SI{0.004}{ppm}, Mn \SI{0.01}{ppm}
  and Co \SI{0.2}{ppm}).
  The Ag film deposited of 99.99 \% pure
  starting material was observed to contain
  higher impurity levels
  (Fe \SI{15}{ppm}, Cr \SI{5}{ppm}, Ni \SI{2}{ppm},
  V \SI{0.01}{ppm}, Mn \SI{20}{ppm} and
  Co \SI{0.08}{ppm}) \cite{eag}.
  The impurities in the source materials
  can also originate from the surface of the
  source material. Materials formed in
  a solution can result in lower
  surface contamination. This type of
  materials are coined "shot". 
  By using
  99.99~\% shot Ag, 99.999 \% shot Ag and 99.999 \% Ag,
  we were able to reduce the level of Fe impurities below
  \SI{3}{ppm}, but not lower. There was no significant
  difference in the observed
  impurity concentration of the
  metal films evaporated of these
  three source materials.



  \section{Summary and outlook}
  
  In conclusion, we have measured thermal
  properties of thin Ag and Cu wires at
  sub-kelvin temperatures and observed
  an anomalously large specific
  heat in the Cu wires.
  Slow thermal relaxation in thin film Cu wires
  was previously reported in
  \cite{Viisanen:2015_relax,Viisanen:2015_calorim},
  whereas here we trace the long time constant to a
  large heat capacity, rather than small
  thermal conductance.
  The Ag samples are observed to follow the free-electron
  estimate, consistent with literature values measured in bulk.
  The band structures of Ag and Cu are similar,
  but the energies of the
  d-electrons are clearly lower in Ag \cite{Ag_Cu_band:1964},
  which can explain why the free-electron model
  applies better for Ag.
  However, the observed anomaly in the electronic specific
  heat of Cu is much larger than that previously measured in bulk,
  exceeding the free-electron estimate by even an order of
  magnitude.
  The indication that the thicker
  wire would have lower specific heat suggests that
  the anomalously high specific heat could be
  related to the surface of the metal.
  However, more experiments are needed
  to confirm this conclusion.
  Magnetic impurities in Cu might increase the
  specific heat of the metal
  at low temperatures.
  At temperatures below $T_\mathrm{K}$,
  the specific heat of an electron gas
  interacting with magnetic impurities
  is given by $c = \kB T/T_{\mathrm{K}}$,
  which can be significantly larger than the
  free electron estimate.
  Around $T_\mathrm{K}$, $c$ develops a
  Kondo peak even at zero magnetic field,
  changing the power
  law of the specific heat\cite{magnetic_impurities:2012}.  
  Yet the concentration of
  known magnetic impurities in the evaporated Cu
  was measured to be as low as \SI{0.6}{ppm} but
  \SI{40}{ppm} in Ag.
  However, the natural Cu oxides, $\mathrm{CuO_{2}}$
  and CuO, are both magnetic,
  being a possible source of Kondo
  impurities in the measured samples.
  Another possible
  source of the enhanced specific heat
  can arise from the quadrupole splitting of the
  nuclei, due to the EFG generated by the
  distortion of
  the lattice\cite{Enss:2000_calorimeter}.
  The lattice of the evaporated Cu is imaged to
  have an irregular shape,
  whereas the Ag lattice is observed to be completely
  uniform. This would suggest the
  quadrupole splitting
  to be more likely to occur in the Cu samples.
  Fabricating much thicker Cu wires as well as wires
  with a passivated surface would provide
  valuable information for distinguishing between
  a surface effect and other possible sources of
  heat capacity.
  Further interesting measurements
  for understanding the
  phenomena now observed would be
  to explore a wider range of
  temperatures and looking at the
  dependence of the heat capacity
  on magnetic field.
  Based on our results, Ag exhibits
  a more promising candidate as
  an absorber material for a nanocalorimeter. \\

%

We acknowledge O.-P. Saira for the collaboration
and discussions in the development of the
RF-NIS thermometer, and A. Peltonen and H. Jiang
for the chemical and
structural analysis of the normal metal wires.
J. Peltonen is acknowledged for contributing to the
fabrication and design of the samples and M. Meschke
for technical support with the measurement devices. We thank L. Wang, C. Enss, F. Hekking, H. Courtois, H. Pothier, J. Ankerhold, J. Ullom, Y. Galperin and M. Krusius for discussions.
We acknowledge Jenny and Antti Wihuri foundation and
Academy of Finland projects 273827 and 285494
for financial support, and facilities and technical support
provided by Otaniemi research infrastructure
for Micro and Nanotechnologies (OtaNano).


	\bibliography{bibliography_2}{}
	\bibliographystyle{apsrev4-1}

\end{document}